# Probing the Intrinsic Properties of Exfoliated Graphene: Raman Spectroscopy of Free-Standing Monolayers

*Stéphane Berciaud[1,2], Sunmin Ryu[2], Louis E. Brus[2], and Tony F. Heinz[1*]*

[1]Departments of Physics and Electrical Engineering, [2]Department of Chemistry

Columbia University, New York, NY 10027, USA

*Email: tony.heinz@columbia.edu

The properties of pristine, free-standing graphene monolayers prepared by mechanical exfoliation of graphite are investigated. The graphene monolayers, suspended over open trenches, are examined by means of spatially resolved Raman spectroscopy of the G-, D-, and 2D-phonon modes. The G-mode phonons exhibit reduced energies (1580 cm$^{-1}$) and increased widths (14 cm$^{-1}$) compared to the response of graphene monolayers supported on the SiO$_2$–covered substrate. From analysis of the G-mode Raman spectra, we deduce that the free-standing graphene monolayers are essentially undoped, with an upper bound of $2 \times 10^{11}$ cm$^{-2}$ for the residual carrier concentration. On the supported regions, significantly higher and spatially inhomogeneous doping is observed. The free-standing graphene monolayers show little local disorder, based on the very weak Raman D-mode response. The two-phonon 2D mode of the free-standing graphene monolayers is downshifted in frequency compared to that of the supported region of the samples and exhibits a narrowed, positively skewed line shape.

Since its recent discovery, graphene has stimulated much experimental and theoretical research[1]. The unusual electronic structure of this two-dimensional material and its distinctive transport properties[2,3] render graphene an intriguing new material for applications such as nanometer-scale field-effect transistors (FETs)[4] and chemical sensors.[5] Although several promising routes have emerged to grow graphene epitaxially[6,7] or to solubilize macroscopic quantities of graphene[8], mechanical exfoliation of graphite[9] currently remains the preferred method to produce highly crystalline graphene samples of both single- and few-layer thickness. Recent electrical and optical characterization of graphene monolayers prepared by mechanical exfoliation has shown that not only processed materials in FET structures,[10-17] but also pristine, unprocessed graphene monolayers deposited on solid surfaces exhibit significant doping.[18] This unintentional doping, which can reach levels of ~$1\times10^{13}$ cm$^{-2}$ under ambient conditions[18], has been found to vary from sample to sample[10,18] and also to be inhomogeneously distributed on a sub-micron scale within a given graphene sample.[17] These doping effects are understood to play a major role in defining the transport properties of typical graphene samples.[10-12] However, the underlying process responsible for doping has not yet been elucidated. In particular, the relative role of doping induced by the substrate and that arising from the intrinsic environmental sensitivity of graphene[19] remains unclear. Free-standing graphene monolayers have recently been produced[20] and shown to exhibit favorable transport properties.[21,22] Such free-standing monolayers provide an excellent reference system in which to address the issue of self-doping effects and to examine the intrinsic properties of this two-dimensional material.

In this Letter, we present results of an investigation of the properties of pristine exfoliated graphene samples that are free from any perturbation of a substrate. This is achieved by preparing graphene monolayers that are suspended over micron-sized trenches. These model graphene samples are characterized by spatially resolved Raman spectroscopy, a purely optical method that can be applied without subjecting the graphene samples to any processing steps. From analysis of the intensity, frequency, and linewidth of the characteristic G-, D-, and 2D-Raman modes, we are able to deduce valuable information about the quality, doping level, and strain exhibited by the samples and, importantly, about the spatial variation of these quantities. We find that free-standing graphene samples prepared under ambient conditions exhibit no evidence of doping effects and a high degree of spatial homogeneity. Specifically, based on analysis of the Raman G-mode response, we obtain an upper bound of $2\times10^{11}$ cm$^{-2}$ for the residual doping level, with no measurable variation across the surface of the graphene samples within our spatial resolution of ~500 nm. The defect density, as reflected by the D-mode strength, was consistently low, and no significant strain could be identified through measurements of polarized G-mode Raman scattering. Comparison of the free-standing region of the graphene monolayers with the portion of the sample supported by the SiO$_2$-covered substrate revealed dramatic differences: the supported areas exhibited high (up to ~$8\times10^{12}$ cm$^{-2}$) and spatially inhomogeneous levels of hole doping. In addition to these findings, based primarily on detailed analysis of the Raman G-mode response, we also report on the characteristics of the 2D mode of the graphene samples. We find that the ratio of the integrated intensities of the 2D and G peaks is systematically lower on the doped, supported regions of the graphene monolayers than on neighboring suspended regions. Interestingly, the 2D mode of free-standing graphene monolayers is also found to exhibit a line shape with significant asymmetry. This contrasts with the broadened, symmetric, and frequency-upshifted feature observed for the 2D mode on the supported regions of the graphene monolayers. Our demonstration of the undoped character of our free-standing graphene monolayers complements recent electrical transport measurements on FETs composed of suspended graphene channels.[21] In these structures, the minimum conductivity occurred near zero gating voltage, which indicates a low intrinsic level doping in the graphene monolayer. This latter study was, however, performed on processed graphene samples that were subsequently annealed at high temperature. The condition of these graphene monolayers might thus be quite different from that of the pristine exfoliated samples analyzed here.

For our investigations we sought a probe that could provide detailed information about the graphene monolayers without the use of electrodes or other structures that would have necessitated processing steps. Raman spectroscopy, as a purely optical technique, meets this requirement and also allows for spatial mapping of the measured characteristics. Raman spectroscopy has been shown to be a very versatile tool for the characterization of graphene.[23] It is sensitive to key material properties, including doping,[18] defect density,[23,24] temperature,[25] and strain.[26] The Raman spectrum of graphene is dominated by three main features, each having a different physical origin. The G mode (at a Raman shift of ~1580 cm$^{-1}$) arises from emission of zone-center optical phonons, whereas the doubly-resonant disorder-induced D mode[27] (~1350 cm$^{-1}$) and the symmetry-allowed 2D (or G') overtone mode (~2700 cm$^{-1}$) involve preferential coupling to transverse optical phonons near the edge of the Brillouin zone.[24, 28-31] The line shape of the 2D peak has been widely used to distinguish single-layer graphene from multi-layered graphene films,[24, 29, 32] while the strength of the D mode is indicative of degree of short-range disorder in the sample. Recent Raman studies on graphene FETs have demonstrated that both the frequency and the linewidth of the G mode can be used to monitor the doping level.[13-17] For electrically neutral graphene, the G-mode frequency reaches its lowest value, while its width is at its maximum. Electron or hole doping leads to a stiffening of the G-mode phonons and an accompanying reduction in its linewidth. The frequency of the 2D mode has been found to shift in opposite directions depending for electron and hole doping. This response can assist in the determination of the sign of doping, at least for significant ($>5\times10^{12}$ cm$^{-2}$) carrier concentrations.[15]

In our study we investigated pristine monolayer graphene samples that were suspended above the substrate. To this end, we prepared a $10\times10$ mm$^2$ array of rectangular trenches (width of 4 µm, length of 86 µm, separated by 16 µm) by etching through the 300-nm SiO$_2$ epilayer that covered our Si substrate (Fig. 1). The patterning was done using conventional ultraviolet photolithography and reactive-ion etching. In order to eliminate possible residues from fabrication, the patterned substrates were annealed in an argon atmosphere at 600 °C for 2 hours and further cleaned using a buffered mixture of sulfuric acid and hydrogen peroxide. Flakes of kish graphite (Toshiba Ceramics) were deposited on the patterned and cleaned substrate under ambient conditions (i.e., in air at 23 °C) by means of the well-established method of mechanical exfoliation.[9] The resulting graphene layers were not subjected to any processing whatsoever and were thus free of potential fabrication-induced modification. The deposited flakes were examined with an optical microscope to identify those graphene samples that were positioned to span etched trenches. Graphene flakes of monolayer thickness were unambiguously identified through Raman measurements of the 2D mode.

Our Raman characterization of these free-standing graphene monolayers was performed under ambient conditions. The Raman spectra were collected in a backscattering geometry using linearly polarized radiation at 514.5 nm from an Ar-ion laser. The laser beam was focused to a spot size of ~500 nm on the graphene samples. We obtained Raman spatial maps by raster scanning with a precision two-dimensional stage having a step size of 500 nm. For such spatial mapping of the Raman response, we generally used a spectral resolution of ~8 cm$^{-1}$ (obtained with a 600 grooves/mm grating); for the measurement of key spectra, however, a spectral resolution of ~2 cm$^{-1}$ (1800 grooves/mm grating) was chosen to elucidate the details of the line shape. In our analysis of linewidths, the G-mode features were fit to a Voigt profile (assuming a Gaussian contribution corresponding to our spectral resolution), while the 2D-mode spectra were fit to Lorentzian profiles. Throughout this letter, the widths $\Gamma_G$ and $\Gamma_{2D}$ of the Raman features are defined, respectively, as the full-width at half maximum (FWHM) of the Lorentzian component of the Voigt profiles and as the FWHM of the Lorentzian fits. All measurements were performed using laser powers of ~700 µW. In order to rule out temperature-induced effects,[25] we compared Raman spectra recorded with laser powers ranging from ~150 µW to ~2 mW. No spectral shifts or changes in the line shape were observed. (See supporting information.)

Figure 1 shows Raman spectra recorded on the supported and free-standing portions of a specific single-layer graphene sample. The Raman spectrum of the suspended part of the graphene monolayer is clearly different from that of supported part. First, the Raman G-mode feature of suspended graphene (1580 cm$^{-1}$) is appreciably downshifted (by 7 cm$^{-1}$) compared with that measured on the supported region. The spectrum obtained from the free-standing graphene also exhibits a much greater width (14 cm$^{-1}$) than that of the supported graphene (6 cm$^{-1}$). Second, the Raman 2D mode of the free-standing graphene is downshifted with respect to the supported portion. Third, the ratio of integrated intensity of the 2D and G features, $I_{2D}/I_G$, is enhanced by a factor of ~4 for the free-standing part of the graphene samples compared with the supported portion. With regard to the D mode, however, there is no significant disorder-induced band on either the suspended or supported regions; the ratio of integrated intensities for the D to G modes is only ~5%.

Some observations may immediately be made based on these results. First, the frequency of the G-mode of the free-standing graphene sample is very similar to that found in single crystals of graphite[33] (~1580 cm$^{-1}$). Also, as we discuss below in more detail, the width of the G-mode of suspended graphene is similar to the highest values recorded for electrostatically gated graphene samples on a substrate,[13-16] a condition that occurs near the point of charge neutrality. This suggests that there is little intrinsic doping in the free-standing graphene monolayers. For the supported portions of graphene sample, we observe a narrowed width and a stiffening of the G-mode phonons, features that indicate substantial doping. From the Raman data we obtain an estimated sheet carrier density of $6\times10^{12}$ cm$^{-2}$, which implies a shift in the Fermi energy of ~250 meV from the Dirac point.[13] With respect to defects, the lack of intensity in the D band indicates that probed regions of both the suspended and supported graphene monolayers have low (and comparable) levels of localized disorder. An effective domain size of ~1 µm can be estimated from the measured D-to-G intensity ratio using the empirical scaling relation established for graphite.[34,35]

In order to obtain a deeper understanding of the properties of the free-standing graphene monolayers and further substantiate the observations given above, Raman spatial mapping was performed on our samples (Figs. 2 and 3). Figure 2 presents spatial images of the frequency $\omega_G$ of the G-mode phonon (Fig. 2a) and its FWHM $\Gamma_G$ (Fig. 2b), as well as the integrated D-to-G intensity ratio $I_D/I_G$ (Fig. 2c). From these results we see that the spectra of Fig. 1 are indeed representative of the behavior at all points of the graphene samples: The suspended portion can always be distinguished from the supported region through its lower value of $\omega_G$ and its enhanced width $\Gamma_G$ for the G mode. The correlation between the frequency $\omega_G$ and its width $\Gamma_G$ for different spatial locations on the suspended and supported portions of the sample is presented in Fig. 2d. On the free-standing region of the graphene monolayer are narrowly, the frequency and width are narrowly distributed around the mean values of $\omega_G$ = 1581 cm$^{-1}$ and $\Gamma_G$ = 13.5 cm$^{-1}$. On the supported portion of the sample the data are more widely scattered, but the overall correlation resembles that previously reported for measurements on several different graphene flakes.[18] Here, however, the inhomogeneous doping occurs for different regions of a single supported graphene monolayer. The physical origin of the correlation of G-mode linewidth and frequency is assumed to be that seen in graphene monolayers under electrostatic gating: the effect of coupling of the phonons to electron-hole pairs, as manifested in the real and imaginary parts of the phonon self-energy.[13-16]

The Raman spatial maps also reveal some distinctive aspects of particular regions of the sample. The most pronounced effect is the altered condition of the upper portion of the supported graphene monolayer. This region exhibits a much greater degree of disorder than all other parts of the sample, whether supported or suspended. Here the intensity of the otherwise weak D mode is as great as 20-100% of that of the G mode. In contrast, elsewhere on the sample (see Fig. 2c), both in the supported and suspended regions, the D-mode intensity is below our detection sensitivity (< 5% of the G-mode intensity under the experimental conditions used to record such maps). The region also appears to be heavily doped, as gauged by the strong upshift in the G-mode frequency. A corresponding reduction in the G-mode linewidth is not, however, observed (grey triangles in Fig. 2d). This presumably reflects the influence of inhomogeneous broadening on the phonon linewidth, an effect arising from variations on a spatial scale below the resolution of our measurement. A further interesting aspect of the Raman maps is the behavior observed at the edges of the graphene sample. On the lower (undamaged) portion of the supported graphene sample, the G mode exhibits an increased frequency and a somewhat reduced linewidth. This effect is particularly evident on the left edge of the sample. We attribute this response to enhanced doping along the edge of the graphene sample. On the other hand, there is little evidence of edge-doping effects on the free-standing portion of the graphene monolayer. Only a slight upshift (of +2 cm$^{-1}$) and reduction in linewidth (of 2 cm$^{-1}$) for the G mode are observed on the left edge, while the right edge shows features similar to those of the central area of the free-standing monolayer.

Figure 3 presents spatial images of related to features of the 2D mode for the same graphene sample. Figures 3a and 3b display the mode frequency $\omega_{2D}$ and linewidth $\Gamma_{2D}$, while Fig. 3d shows the correlation between these two parameters. We see that the 2D feature on the supported portion of the sample exhibits a greater width (about +5 cm$^{-1}$) and is significantly upshifted in frequency (about +10 cm$^{-1}$) compared to the response of the free-standing graphene monolayer. The tip of the sample, which has previously been identified as a highly doped and disordered region, shows the highest 2D-mode frequency and is associated with increased and scattered values of the linewidth. The overall stiffening of the 2D mode in the highly-doped, supported regions allows us to conclude that supported graphene is hole doped.[15]

The ratio of the integrated intensity of the 2D mode compared to that of the G mode, $I_{2D}/I_G$, is presented in the spatial map of Fig 3c. On the suspended portion of the sample, the 2D-mode intensity is enhanced, while the G-mode strength is not strongly affected. Consistent with studies of top-gated graphene FETs,[15] as well as with recent theoretical results suggesting a reduced intensity of the doubly resonant features in doped graphene from enhanced electron-electron scattering,[36] this ratio provides a criterion to distinguish neutral and doped graphene. We note that the absolute intensities for G- and 2D-modes measured in the free-standing and supported regions might differ simply because of an optical etalon effect associated with the different dielectric environments. However, the ratio of two mode intensities for the different regions, $I_{2D}/I_G$, will hardly be changed with the environment, since the wavelengths of the scattered photons (560 nm and 597 nm) are so similar.[37]

We also recorded Raman maps on bilayer graphene samples. Such samples exhibited qualitatively similar features, but with less pronounced differences between the free-standing and supported portions (see supporting information). In particular, a decrease in the effective charge transfer from molecular adsorbates,[19] as well as the smaller shift in the Fermi energy for a given doping level in bilayer graphene,[38] may account for the reduction in the G-mode stiffening observed for supported bilayers.

Table 1 presents a summary of Raman data for the G- and 2D-modes for three single-layer graphene samples that included both a free-standing and supported region. The indicated mean values (and standard deviations) of the frequencies ($\omega_G$, $\omega_{2D}$) and linewidths ($\Gamma_G$, $\Gamma_{2D}$) have been obtained from analysis of the complete Raman spatial maps of the samples. The results are consistent with those shown in Figs. 1-3 which were obtained from sample A). In discussing these results, we first consider the behavior of the G mode. For the free-standing regions of the three samples, little variation is seen in the mean values of the G-mode frequency and linewidth: The average values of $\omega_G$ spans a range of only 3 cm$^{-1}$, while the average value of $\Gamma_G$ for the three samples varies by just 0.5 cm$^{-1}$. These results highlight the degree of consistency and spatial homogeneity of the free-standing graphene monolayers. In contrast, the supported portions of the samples exhibit greater inhomogeneity: For the three samples, the average values of $\omega_G$ vary by 9 cm$^{-1}$ and the mean widths $\Gamma_G$ differ by 4.5 cm$^{-1}$. Similarly, the variation of these parameters *within* the supported region of each sample exceeds that for the suspended region, as indicated by the measured standard deviations for each of these quantities shown in Table 1.

We now turn to a more quantitative discussion about the implication of our data for the doping levels in the pristine, free-standing graphene monolayers. Such an analysis can be based either on consideration of the frequency or the linewidth of the Raman G mode, since both of these quantities are affected by the doping level of the graphene through the existence of strong electron-phonon coupling[39-41]. The frequency of the G mode is often used as a measure of the doping level.[18, 19] For low temperatures, $\omega_G$ should exhibit a (somewhat complex) dependence on doping concentration, even at low levels. At room temperature, however, this effect is washed out; the G-mode frequency is expected to reach an essentially constant minimum value for all carrier densities below ~1×10$^{12}$ cm$^{-2}$.[39] In this context, the low observed peak frequency $\omega_G$, with little dispersion around its mean value, is consistent with a low level of doping of suspended graphene. However, it does not provide a stringent quantitative estimate of the residual carrier concentration.

For our room-temperature studies, we find the linewidth of the G-mode feature to be a more sensitive probe of low doping levels. Indeed, considering the broadening of the G mode to be proportional to the statistical availability for electron-hole pair generation at the G-mode energy,[39] we can write the G-mode linewidth as

$$\Gamma_G = \Gamma_0 + \Delta\Gamma \left[ f_T(-\hbar\omega_G/2 - E_F) - f_T(\hbar\omega_G/2 - E_F) \right] \quad . \qquad (1)$$

Here $\Delta\Gamma$ denotes the maximum phonon broadening from electron-hole pair generation (Landau damping) as it would occur at zero temperature; $f_T$ is the Fermi-Dirac distribution at temperature $T$; $E_F$ is the Fermi energy relative to the Dirac point in graphene; and $\Gamma_0$ is the contribution to the linewidth from phonon-phonon coupling and other sources that are independent of these electronic interactions. At room temperature, the maximum contribution to the broadening from the electron-hole pair coupling is ~0.95 $\Delta\Gamma$ and will be approximated as $\Delta\Gamma$ in the following discussion.

In other studies, the maximum values of $\Gamma_G$ observed for graphene FETs range from 14 cm$^{-1}$ to 16 cm$^{-1}$; however, all of these investigations agree on a value of $\Delta\Gamma$ of 9±0.5 cm$^{-1}$ for the electronically induced component of the linewidth.[13-16] Our experiment did not permit us to calibrate the Raman measurements with a systematic variation of doping, since no gate was present in our sample geometry. However, the lowest value of $\Gamma_G$ observed on the strongly doped supported regions of sample A (5±0.5 cm$^{-1}$, see Fig. 2d) provides an upper bound for the intrinsic (non-electronic) linewidth $\Gamma_0$. The difference between this estimate of $\Gamma_0$ and the mean value of $\Gamma_G$ in suspended graphene (14±0.5 cm$^{-1}$) is in excellent agreement with the previously reported values of $\Delta\Gamma$. Within our experimental accuracy, we then deduce from (1) that the average doping level of our suspended graphene samples does not exceed $2\times10^{11}$ cm$^{-2}$ (see supporting information for details on this estimate). In addition, the small standard deviation around the mean values of $\Gamma_G$ (< 1cm$^{-1}$, see Table 1) implies that the inhomogeneity of the doping level of the suspended graphene monolayers is less than $2\times10^{11}$ cm$^{-2}$ (when averaged over the 500 nm spot size of the probe laser beam). The data correspond to an aggregate scanned area of at least 30 µm$^2$ (see samples B, C in Table 1 and also the supporting information). This result confirms the homogeneity of the low doping level of unprocessed, free-standing graphene monolayers.

Our interpretation of the features of the Raman spectra presented above has neglected the possible influence of strain on the graphene samples. Here we argue that strain effects cannot be significant. Given our sample geometry for the suspended samples, the existence of isotropic strain can reasonably be excluded. However, residual uniaxial tensile stress, perpendicular to the trench, might be present and could result in a downshift of the G mode. This would complicate our interpretation of the frequency shifts in terms of doping level. Further, the presence of anisotropic strain might produce an effective line broadening. To test such a scenario experimentally, we carefully examined the polarization dependence of the Raman spectra of the suspended portions of our samples. We recorded the G-mode response for different detected polarizations of the Raman scattered light under fixed linear polarization of the pump beam (see supporting information). Within our experimental accuracy with 2 cm$^{-1}$ spectral resolution, we did not observe any measurable shift (< 0.5 cm$^{-1}$) or modification of the line shape. Since anisotropic strain would lead to a splitting of the Raman G modes, this null result confirms that the G mode of suspended graphene is not significantly altered by mechanical stress. More precisely, from the experimentally determined shift rates of the high- and low-frequency modes (-5.6 cm$^{-1}$ per % strain and -12.5 cm$^{-1}$ per % strain, respectively),[26] we can estimate the residual strain in the free-standing graphene samples to be less than 0.1%.

We now consider the behavior of the 2D band for the free-standing graphene films. We first comment on its frequency and then discuss its width and line shape. As shown in Table 1, the frequency $\omega_{2D}$ for the free-standing graphene monolayers, near 2674 cm$^{-1}$ for all three samples, is lower than that found for the supported portions of the graphene films. Interestingly, the supported part of sample B shows only a weak average upshift (+1 cm$^{-1}$) with respect to the free-standing region. For sample C, on the other hand, a significant stiffening of the 2D band (+4.5 cm$^{-1}$) is observed on the supported part, despite its low estimated doping level of ~5×10$^{11}$ cm$^{-2}$ (from the G-mode linewidth). This comparison suggests that doping alone cannot explain the different behavior of the 2D mode observed on our samples. Because of the doubly-resonant nature of the 2D mode, the measured response reflects both the electron and phonon structure of graphene.[27-31] Thus changes in either the electron or phonon dispersion relations may contribute to shifts in the 2D frequency.

On the free-standing graphene monolayers, the average width $\Gamma_{2D}$ of the 2D mode is 23 cm$^{-1}$, with little deviation from its central value seen in the different samples. This width is significantly narrower than that measured on supported graphene (see Table 1). This observation is surprising, since no significant change in the 2D linewidth has been reported as a function of doping level for electrostatically gated samples.[13, 15-17] In addition to the reduced linewidth for free-standing graphene compared with supported graphene, we also find a marked difference in the line shape. The Raman spectra in Fig. 4, taken in the high-resolution configuration, compare the line shape of the 2D mode on the free-standing and supported region of a sample. For the supported graphene monolayers, the 2D peak is symmetric and fits a Lorentzian profile. On the other hand for the free-standing graphene monolayer, the 2D feature is positively skewed. The 2D peak has a narrower half width on the low-frequency side, while its high-frequency wing has the same line shape as that for supported graphene. The origin of this distinctive line shape for the 2D mode of suspended graphene is not currently understood. A full two-dimensional calculation of the Raman response of the 2D, taking into account trigonal warping effects as well as the details of the phonon dispersion near the K point,[42] would be highly desirable.

In conclusion, we have examined, using spatially resolved Raman scattering, the properties of pristine, free-standing graphene monolayers prepared by mechanical exfoliation. The samples were prepared under ambient conditions without any processing. These graphene monolayers are found to be spatially homogeneous and to differ significantly from the properties of the same samples supported on an oxide-coated silicon substrate. In particular, the free-standing graphene monolayers showed no intrinsic doping, with an upper bound for the residual carrier density of $2 \times 10^{11}$ cm$^{-2}$, while the supported portions of the graphene exhibited spatially varying doping levels up to ~$8 \times 10^{12}$ cm$^{-2}$. The density of defects, as measured by the Raman D mode, lay below our experimental sensitivity for the suspended graphene samples. We conclude that the exfoliation process itself and the associated exposure to ambient atmospheric conditions do not lead to surface chemical reactions that significantly perturb the graphene lattice structure or cause appreciable charge transfer to the graphene layer. Thus the (spatially inhomogeneous) doping observed here and reported in the literature for supported graphene samples arises from the interaction of the graphene with the substrate, rather than as an intrinsic property of exfoliated graphene prepared and held under ambient conditions. This finding implies that improvements in the charge transport properties of graphene may be possible through modification and better control of the substrate.

**Acknowledgement.** We would like to thank Changgu Lee, Sami Rosenblatt, and Hugen Yan for helpful discussions. We acknowledge support from the Nanoscale Science and Engineering Initiative of the NSF (grant CHE-0117752), the New York State Office of Science, Technology, and Academic Research (NYSTAR), the Nanoelectronics Research Initiative, and the Office of Basic Energy Sciences, U.S. DOE (grants DE-FG02-98ER-14861 and DE-FG02-03ER15463).

**Supporting Information Available:** Raman spectra as a function of the laser power, Raman maps of suspended samples of mono- and bilayer graphene, estimation of the residual doping level, and results of Raman spectra obtained for different detected polarizations. This material is available free of charge via the internet at http://pubs.acs.org

**Figures and Captions.**

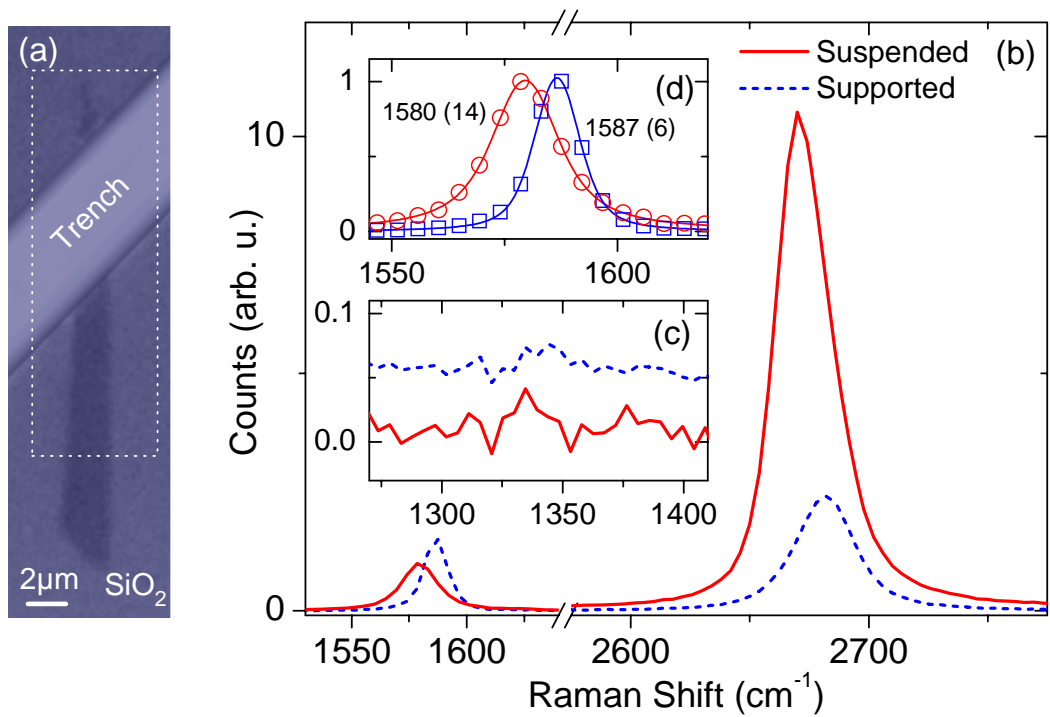

**Figure 1**: (a) Optical micrograph of an exfoliated graphene monolayer spanning a 300-nm deep trench etched in the $SiO_2$ epilayer. The supported region shows appreciable optical contrast, while the free-standing graphene layer cannot be seen in the image. (b) Raman spectra recorded on the single-layer graphene sample of (a), both for the suspended (red solid line) and supported regions (blue dashed line). (c) Raman spectra in the lower-frequency region (1270-1420 cm$^{-1}$) for the two regions of the sample. The curves are offset for clarity. The spectra in (b) and (c) have been normalized with respect to the integrated intensity of the G mode. (d) Detailed comparison of normalized spectra for the G mode. The experimental data (red open circles and blue open squares, for suspended and supported graphene, respectively) are fit with Voigt profiles (solid lines). The energy (width) of the G mode is indicated in cm$^{-1}$ for each spectrum.

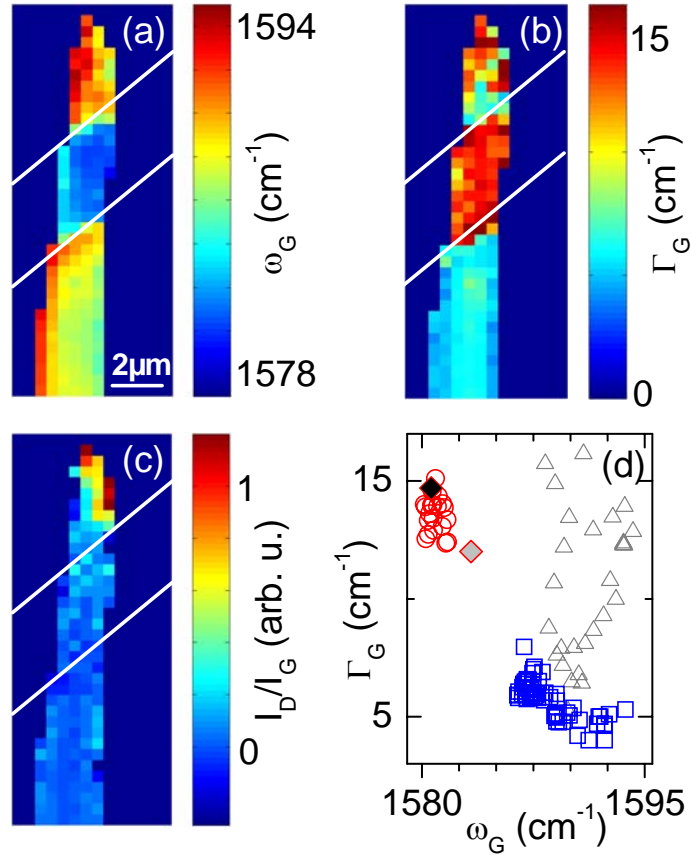

**Figure 2**: Spatial maps of the Raman features of a single-layer graphene sample, with regions of free-standing and supported material: (a) the G-mode frequency, $\omega_G$; (b) the G-mode linewidth, $\Gamma_G$; and (c) the ratio of the integrated intensity of the D mode to that of the G mode, $I_D/I_G$. The data were recorded over the boxed region of the graphene sample of Fig. 1a. The white lines designate the borders of the trench. Data in (a) and (b) are based on fits of a Voigt line shape to the experimental data. (d) shows the correlation between $\omega_G$ and $\Gamma_G$ for the suspended (red open circles) and lower supported (blue open squares) regions, as well as for the defective tip of the sample (grey open triangles). The average values measured on the left and right edges of the suspended area are represented, respectively, by the grey and black diamonds.

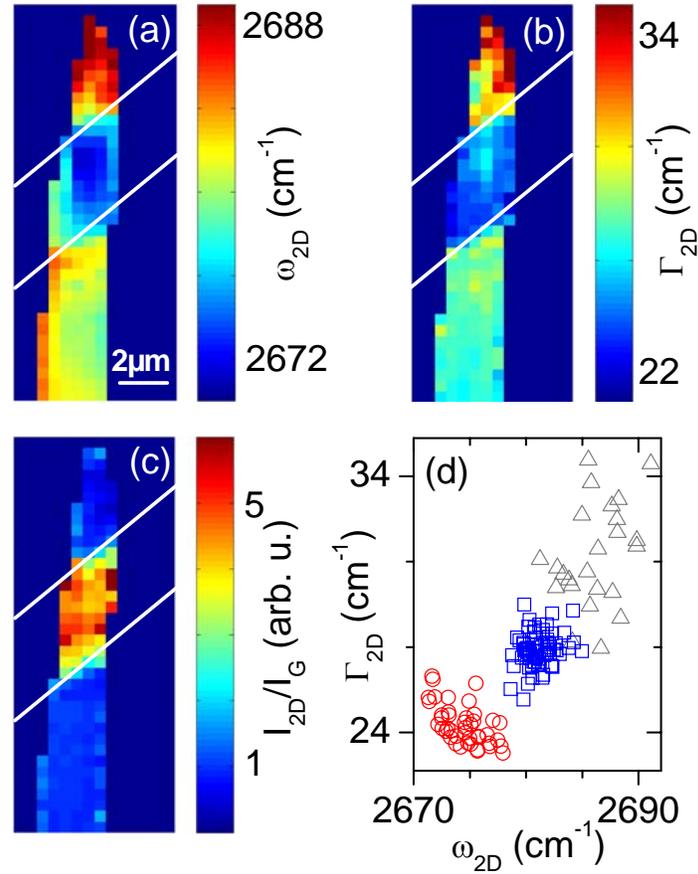

**Figure 3**: Spatial maps for the same sample as in Fig. 2, but presenting results for the Raman spectra of the 2D mode: (a) the mode frequency, $\omega_{2D}$; (b) the linewidth (FWHM), $\Gamma_{2D}$; and (c) the ratio of the integrated intensity of 2D mode to that of the G mode, $I_{2D}/I_G$. The data in (a) and (b) are deduced from Lorentzian fits. The results in (c) are normalized with respect to the average value measured on the lower supported area. (d) shows the correlation between $\omega_{2D}$ and $\Gamma_{2D}$ for the suspended (red open circles) and lower supported (blue open squares) regions, as well as for the defective tip of the sample (grey open triangles).

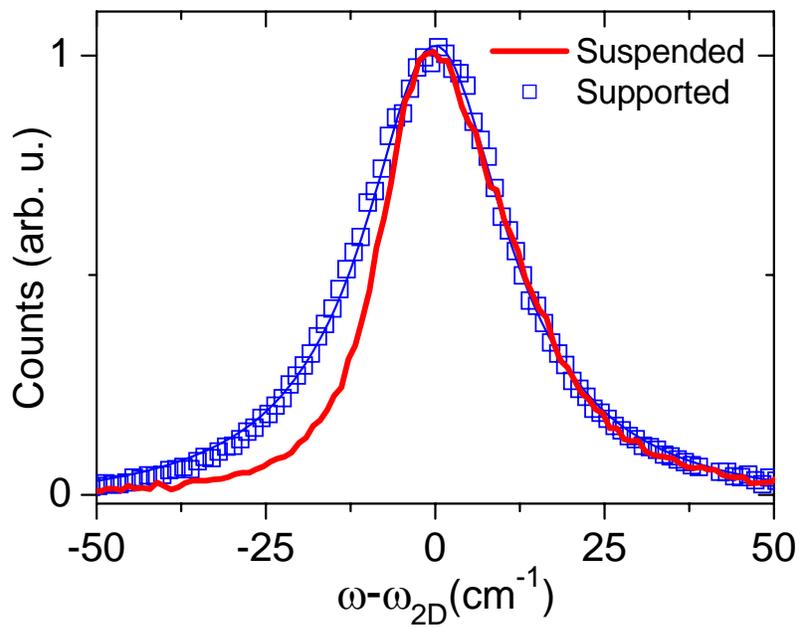

**Figure 4:** Raman spectra for the 2D mode for the supported (open blue squares) and free-standing (thick red line) portion of a graphene monolayer. The two spectra are shifted in energy for a clear comparison. The thin blue line is a Lorentzian fit to experimental data measured on the supported region of the graphene monolayer.

|  | Sample | $\omega_G$ (cm$^{-1}$) | $\Gamma_G$ (cm$^{-1}$) | $\omega_{2D}$ (cm$^{-1}$) | $\Gamma_{2D}$ (cm$^{-1}$) |
|---|---|---|---|---|---|
| A | sus. (7µm$^2$) | 1581.0 ± 0.5 | 13.5 ± 1.0 | 2674.5 ± 2.0 | 24.0 ± 1.0 |
| A | sup. (15µm$^2$) | 1588.0 ± 2.0 | 6.0 ± 1.0 | 2681.0 ± 1.5 | 27.0 ± 1.0 |
| B | sus. (30µm$^2$) | 1579.5 ± 0.5 | 14.0 ± 1.0 | 2673.0 ± 2.5 | 23.5 ± 0.5 |
| B | sup. (40µm$^2$) | 1583.0 ± 1.0 | 8.0 ± 1.5 | 2674.0 ± 2.0 | 27.0 ± 1.0 |
| C | sus. (30µm$^2$) | 1578.0 ± 0.5 | 13.5 ± 0.5 | 2673.0 ± 0.5 | 22.5 ± 0.5 |
| C | sup. (75µm$^2$) | 1579.0 ± 1.0 | 10.5 ± 0.5 | 2677.5 ± 1.5 | 24.0 ± 0.5 |

**Table 1.** Mean values of frequencies and linewidths of the G and 2D modes as deduced from analysis of the spatial maps of the Raman spectra taken on three selected graphene monolayers The aggregate areas of the suspended (sus.) and supported (sup.) portions of the three samples (A-C) are indicated. The uncertainties are standard deviations computed from the spatial maps of the corresponding quantity.